\documentstyle[12pt,fleqn]{article}
\def\be{\begin{equation}}
\def\ee{\end{equation}}
\def\ba{\begin{eqnarray}}
\def\ea{\end{eqnarray}}

\def\fun#1#2{\lower3.6pt\vbox{\baselineskip0pt\lineskip.9pt

\ialign{$\mathsurround=0pt#1\hfill##\hfil$\crcr#2\crcr\sim\crcr}}}\def\re#1{{$
^{\ref{#1}}$}}
\begin{document}
\begin{titlepage}
\null\vspace{-62pt}
\begin{flushright}DOE-ER/40682-33\\
      CMU--HEP93--08\\
      PITT--93--04\\
      April, 1993

\end{flushright}
\vspace{0.3in}
\centerline{{\large \bf Quantum Statistical Metastability Revisited}}
\vspace{0.4in}
\centerline{Daniel Boyanovsky,$^{(a)}$\ \ Richard Holman,$^{(b)}$}

\vspace{0.2in}
\centerline{Da-Shin Lee,$^{(a)}$\ \  Jo\~{a}o P. Silva,$^{(b)}$\ \
Anupam Singh $^{(b)}$}
\vspace{0.3in}
\centerline{{\it $^{(a)}$Department of Physics and Astronomy,
University of Pittsburgh, Pittsburgh PA~~15260}}
\vspace{.2in}
\centerline{{\it $^{(b)}$Physics Department, Carnegie Mellon University,
Pittsburgh, PA~~ 15213}}
\vspace{.5in}
\baselineskip=24pt
\centerline{\bf Abstract}
\begin{quotation}
We calculate the decay rate for a state prepared in a thermal density matrix
centered on a metastable ground state. We find a rate that is intrinsically
time {\it dependent}, as opposed to the {\it constant} rates of previous works.
The rate vanishes at early times, rises to a maximum and eventually falls-off
to zero as a consequence of unitary time evolution. Finally, we discuss
extensions of this calculation to field theories and possible implications for
both sphaleron mediated transitions and first order inflationary theories.
\end{quotation}
\end{titlepage}
\newpage

\baselineskip=24pt

The analysis of the decay of metastable states has always been a important and
interesting topic in physics. Recently, however, this subject has assumed even
greater importance due to the discovery that there are field configurations in
the standard model (so-called sphalerons\re{sphaleron}) that mediate baryon
number violating transitions which are {\it unsuppressed} at high temperature.
Needless to say, this has important implications for the evolution of baryon
number in the early universe\re{baryogenesis}. Another reason why there has
been a rekindling of interest in the evolution of metastable states in the
early universe is due to the development of viable models of inflation
(extended inflation\re{lastein}) that go back to Guth's\re{guth} idea of ending
the inflationary era via false vacuum decay. In short, there are good reasons
for making sure that the decay of metastable states at finite
temperature is, in
fact, well understood.

We will argue in this Letter that some aspects of the calculation of the decay
rate of metastable states must be rethought. In particular, we will make the
point that different choices of initial state can make for significant changes
in the decay rate. We show that under realistic conditions, the decay rate
{\it must} be time dependent. This is consistent with some recent experimental
data which we discuss below.

Two of the seminal works on this topic are those by Langer\re{langer} and
Affleck\re{affleck}. It will be instructive to review these calculations, since
our results are quite different from theirs.

Langer develops a Fokker-Planck type equation for the probability of finding
the system in a given configuration at time $t$. This probability obeys a
continuity equation and the associated current gives the flow of probability in
the configuration space. In a one dimensional system, this current, evaluated
at the saddle point (which is the top of the barrier), is identified with the
rate of activation of the system over the barrier.

To compute this current, Langer then constructs a {\it steady state} solution
to his Fokker-Planck equation. This is tantamount to setting up a steady state
situation by continously replenishing the metastable state at a rate equal to
the rate at which it is leaking across the activation energy barrier.
This point is emphasized in Langer's work.

In Affleck's calculation, the rate is defined as the Boltzmann average of the
probability current over a set of quantum states that, for energies less than
the barrier height, are standing waves in the metastable well. For energies
higher than the barrier height, these are waves incident from the left,
reflected and transmitted at the barrier. The rate is then calculated as:

\be
\Gamma = Z_0^{-1} \int_{0}^{\infty}\ dE \rho(E) \Gamma(E) \exp(-\beta\ E),
\ee
where $\rho(E)$ is the density of states at energy $E$, $\Gamma(E)$ the decay
rate for states with energy $E$, $\beta = 1\slash k_B T$. The rate is
normalized using the partition function $Z_0$ of a harmonic oscillator centered
at the metastable state at $x=x_0$ and whose frequency is just $\omega_0^2
=V''(x_0)$. As in Langer's calculation, this corresponds to a steady flow of
particles across the barrier.

Both of these calculations (and those that have built on them) assume that the
state under consideration is one in which probability is being fed continously
into the metastable well in order to replenish the probability that is flowing
out and over the activation barrier, thus ensuring a steady state. However, we
would argue that this is {\it not} a realistic initial state to use to compute
a rate for processes relevant in the early universe, though there may be other
physical situations for which such an initial state is appropriate.

Usually in inflationary models, say, one thinks of the field that will drive
inflation as being ``trapped'' in the metastable minimum. This trapping arose
due to the fact that the field was in thermal equilibrium with the ambient
heat bath, and as the temperature dependent effective potential changed shape,
the global minimum became separated from the local one by a barrier. We would
thus expect that the appropriate description for the initial state of the field
would correspond to a {\it thermal density matrix}, centered at the metastable
minimum. As long as the field remained in good thermal contact with the heat
bath, the thermal character of the initial density matrix would be maintained.

This is a very different initial state from that considered by Langer and
Affleck. In particular, there is {\it no} replenishing of probability by an
outside source. Thus we expect the decay rate out of the metastable state to
behave very differently than that found by Langer and Affleck. The rate thus
obtained will be an {\it intrinsically} non-equilibrium quantity as the initial
density matrix will evolve in time. It is to this calculation that we now turn
to.

The procedure is straightforward. Consider a one dimensional quantum
mechanical system for simplicity. Start with an initial density matrix
$\rho(t=0)$ and then evolve it in time via either the Liouville equation:

\be
\label{eq:liouville}
i \hbar \frac{\partial \rho(t)}{\partial t} = [H,\rho(t)],
\ee
or via the solution to this equation:

\be
\rho(t) = \exp(-\frac{i}{\hbar} H t)\rho(0)\exp(\frac{i}{\hbar} H t).
\ee
Here $H$ is the Hamiltonian of the system: $H = p^2/2 + V(x)$, where we choose
units so that the mass $m=1$.

Given the density matrix as a function of time, we can look at its position
space representation $\rho(x,x';t)\equiv \langle x|\rho(t)|x'\rangle$. The
current is then found via:

\be
\label{eq: CURRENT}
J(x,t) = \frac{\hbar}{2 i}(\frac{\partial}{\partial x}-
\frac{\partial}{\partial x'})
\rho(x,x';t)|_{x=x'}.
\ee
Evaluating this current at the saddle point will then give us the rate of
activation over the barrier.

While simple conceptually, an exact calculation following this prescription is
beyond our capabilities. This calculation would require the knowledge of how to
specify the exact initial thermal state given an arbitrary potential $V(x)$, as
well as of how to compute the required propagators. To circumvent these
problems, we will use the following techniques. First, we will assume that for
$t<0$, the potential is just a quadratic centered around the metastable state
at $x=x_0$. This then allows us to compute the initial density matrix
as\re{feynman}:

\ba
\label{eq:initdensmat}
\rho(x,x';t=0) & = & N(0) \exp\left\{-\frac{\omega}
{2\hbar \sinh(\beta \hbar \omega)}
\left[((x-x_0)^2+(x'-x_0)^2)\cosh(\beta \hbar \omega) \right. \right.
 \nonumber \\
               &   & \left. \left.-2(x-x_0)(x'-x_0)\right] \right\}
\ea
where the normalization factor
 $N(0)=\sqrt{{\omega\tanh(\beta \hbar \omega)}\slash{\pi \hbar}}$ ensures
that ${\rm Tr}(\rho(t=0))=1$. This was also the approach followed implicitly by
Langer and Affleck when they normalized their rates.

We then use the sudden approximation and say that for $t>0$, the potential
consists of the barrier whose peak is at $x=0$ (see fig. 1). Furthermore, since
we are only interested in {\it over} the barrier activation, we use the
quadratic approximation $V(x) \simeq V_0 - 1/2\ \Omega^2 x^2$ for the
potential,
where $\Omega^2 \equiv -V''(x=0)$ is positive, (since $x=0$ is a maximum of
$V(x)$) and $V_0$ is the height of the barrier.

The coordinate space expression for the time evolved density matrix is, for
$t>0$:

\be
\label{eq: DENSITY}
\rho(x,x';t) = \int\ dy \ dy' \langle x|\exp(-\frac{i}{\hbar}H_{q} t)|y\rangle
\rho(y,y';t=0) \langle y'|\exp(\frac{i}{\hbar}H_{q} t)|x'\rangle
\ee
where the effective Hamiltonian $H_q$ is the quadratic approximation to $H$
near the top of the barrier: $H_q = p^2/2 + V_0 -1/2\ \Omega^2 x^2$.

The propagators $\langle x|\exp(\pm \frac{i}{\hbar}H_{q} t)|y\rangle$ are easy
to evaluate by analytically continuing the propagator for a standard harmonic
oscillator with real frequency\re{barton}:

\be
\langle x|\exp(\pm \frac{i}{\hbar}H_{q} t)|y\rangle = M(t)
\exp(\pm \frac{i}{2 \hbar} \frac{\Omega}{\sinh(\Omega t)}\left[(x^2+y^2)
\cosh(\Omega t)-2xy\right]).
\ee
with $M(t) =  ({\pm \Omega}\slash {2 \pi i \hbar \sinh(\Omega t)})^{1/2}$.
These propagators are solutions to the evolution equation with the proper
boundary conditions. We can now compute the density matrix as a function of
time, as well as the current. We have verified that the resulting density
matrix is a solution of the Liouville equation (Eq.~(\ref{eq:liouville})) with
the initial boundary condition given by Eq.~(\ref{eq:initdensmat}), thus
confirming that the analytically continued propagators give the correct answer.

Rather than write down the density matrix, we consider the probability
density $p(x, t) \equiv \rho(x, x;t)$:

\be
\label{eq:probdens}
p(x, t)  =  \frac{1}{\sqrt{2 \pi}\sigma (t)} \exp(-(x-x_0 \cosh(\Omega t))^2
\slash {2 \sigma (t)^2})
\ee
with

\be
2\sigma (t)^2  =  \frac{\hbar}{\omega \tanh(\beta \hbar \omega/2)}
(\cosh^{2}(\Omega t) + \frac{\omega^2}{\Omega^2}\sinh^{2}(\Omega t)).
\ee
The rate $\Gamma(t) = J(x=0, t)$ is found to be:

\be
\Gamma(t) = \frac{\omega^2}{\Omega} \left[\frac{\omega}{\pi \hbar}
\tanh(\beta \hbar \omega/2)\right]^{1/2} |x_0| A(t)
\exp\left[-\tanh(\beta \hbar \omega/2) B(t)\right] \label{rate}
\ee
with
\ba
A(t) & = &  \frac{\sinh(\Omega t)}{\left[\cosh^2 (\Omega t) +
\frac{\omega^2}{\Omega^2} \sinh^2 (\Omega t)\right]^{3/2}}\label{Aoft} \\
B(t) & = & \frac{\omega x_0^2}{\hbar}
\left[1 + \frac{\omega^2}{\Omega^2}\tanh^2 (\Omega t)\right]^{-1}
\label{Boft}
\ea

The first feature we should remark on is that both the probability density and
the rate are time {\it dependent}! This is in marked contrast to
both Affleck and Langer's results where the rate had the generic form:
$\Gamma =A\exp(-B)$, with both $A$ and $B$ time {\it independent}. However, it
is easy enough to argue that given our initial state, and the fact that the
transition from the metastable state to the true ground state {\it must} be a
{\it non-equilibrium} process, this time dependence was inevitable.

As the system activates over the barrier, there is loss of probability in the
metastable well. Thus, as time goes on, we should expect to have less and less
``initial state'' to decay from. This is a consequence of unitary time
evolution for the density matrix, and hence for the probability distribution.
{}From this argument, we expect the rate to start at zero initially, and then
rise to some maximum. After this, the rate should then decrease to zero
at large
times. This is exactly the behavior demonstrated by the rate we have calculated
(fig.2).

There are two competing effects that determine the behavior of
the rate: the motion of the center of the probability distribution
(Eq.~(\ref{eq:probdens})) and the spread, determined by $\sigma(t)$. The most
important factor turns out to be the spread. This may be understood from
the fact that the fluctuation $\langle (x-x_0(t))^2 \rangle =
{\sigma^2(t)}$, where $x_0(t) = x_0 \cosh(\Omega t)$.

A remarkable result has been recently reported by Min and
Goldburg\re{mingoldburg} concerning nucleation in a classical fluid under
shear. They found a {\it time dependent} nucleation rate for this system that
is strikingly similar in form to the rate obtained above and strongly dependent
on the initial state. Though our quantum mechanical calculation does not apply
to the classical fluid case, these results show that there are systems in which
the nucleation rate is time dependent. These results {\it cannot} be explained
by the steady-state homogeneous nucleation theory. While our quantum mechanical
calculation only describes one degree of freedom, it may provide qualitative
insight into the dynamics of the collective coordinate that describes the
radius of a droplet. Thus the classical limit of our calculation may still
provide a qualitative description of time dependent rates in macroscopic
situations.

In previous calculations of an activation rate, a Boltzmann supression factor
of the form $\exp(-\beta V_0)$ usually appears. Our result above does not seem
to have such a factor in it. However, a suppression factor of this type does in
fact appear in our $\Gamma(t)$. It is encoded in the relationship between
$\omega$ and $\Omega$. To see this, we need to consider a specific potential
$V(x)$. Thus consider a cubic potential:

\be
V(x) = V_0 (1 + 2 x^3/{x_0}^3 - 3 x^2/{x_0}^2)
\ee
where we take $x_0<0$. With this potential we have $\omega^2=\Omega^2= 6
V_0/{x_0}^2$. If we now take the high temperature limit, we find that the
exponential in the current becomes $\exp(-3\beta V_0 f(t))$, with $f(t) = 1/2 \
(1 + {\rm sech}(2 \Omega t))$; note that $f(t)$ varies from $1$ to $1/2$ as $t$
varies from $0$ to $\infty$. Thus, we do get a suppresion of the Boltzmann
form, but it is larger than $\exp(-\beta V_0)$. This extra
suppression comes about due to the initial state we are using; since it is
centered around $x = x_0$, only a fraction of the probability is near the
saddle point at $x=0$. In fact, $p(x=0,t) \propto \exp(-3\beta V_0 f(t))$ in
the high $T$ limit. The extra suppression is then seen as a measure of how much
(or how little!) support $p(x, t)$ has near $x=0$.

The next feature we examine is the temperature dependence of the rate. The
prefactor in $\Gamma(t)$ contains the factor $\sqrt{\tanh(\beta \hbar
\omega/2)}$ and so decreases at high enough temperature. However, the
exponential is given by $\exp(-\tanh(\beta \hbar \omega/2) B(t))$, which {\it
increases} with temperature. Thus, there will be a temperature regime in which
the rate increases with temperature. In the high $T$ limit i.e. $\beta \hbar
\omega<<1$, we can find this regime by computing $\partial \ln \Gamma\slash
\partial \beta$:

\be
\partial \ln \Gamma\slash \partial \beta = \frac{1}{2\beta}
(1-\beta \hbar \omega B(t))
\ee
Thus, $\Gamma$ increases with $T$ as long as $1-\beta \hbar \omega B(t)<0$. For
the case of the cubic potential above, this becomes $6 \beta V_0>1+\tanh^2
(\Omega t)$. Now, in order to have a metastable state at all, we should require
that the temperature be less than the barrier height $V_0$. Thus, to the extent
that we have an initial state that can be thought of as trapped in the
metastable well (which is to say, $\beta V_0>1$), the rate will increase with
temperature (fig.3). This requires $V_0>>\hbar \omega$ as a consistency
condition. This is also a required condition for the initial state to be
thought
of as metastable.

A valid point to raise at this time concerns the validity of our
approximations. Certainly, if we quench the system quickly enough, it will
settle into the metastable well. Furthermore, as long as $1\slash 2\ \hbar
\omega<V_0$, we expect the harmonic oscillator approximation for
the potential near $x=x_0$ to be reasonable.
While our calculations were made within the sudden approximation, the fact that
the density matrix and hence the rate will be time dependent will still remain
even under more realistic time evolution. The reason for this is, again, the
fact that the density matrix evolves in a unitary way.

The last truncation of the original theory was the use of the quadratic
approximation for the potential near the top of the barrier. This gives rise to
the motion of the center of the probability distribution as well as its spread
as described in eq.(\ref{eq:probdens}). After a time $\tau \sim \Omega^{-1}$,
however, we expect the effects of the non-quadratic terms of the potential to
make themselves felt and modify our result. Thus, our expression for
$\Gamma(t)$
can be trusted for time of order $\Omega^{-1}$. As can be seen from,
Figure 2, however, this is sufficient time to see the rise of $\Gamma(t)$ to
its peak value and to see the beginning of its decrease to zero. The fact that
$\Gamma(t)$ will approach zero asymptotically is, as mentioned previously, just
a consequence of unitary time evolution of $\rho(t)$ and hence can be trusted.
However, the exact shape of the curve may be different than that shown
in Figure 2.

We started this work by using examples such as first order inflationary models
and baryon number violation via sphaleron mediated decays to motivate the
discussion. What does our calculation say about these topics? In order to
truly extract some information, we need to understand how our
calculation should
be generalized within the context of field theory. Some steps in this
direction have already been taken. Boyanovsky and Arag\~{a}o de Carvalho\re{dc}
have considered the problem of thermal activation over a barrier in a $1+1$
dimensional scalar field theory. While their calculation involves some
sublteties not present in the quantum mechanical case (such as dealing with
collective coordinates), the results are similar. They arrive at a rate that
varies in time much like the rate we find here. We are also currently involved
in calculations of decay rates in theories involving sphalerons using the
real-time formalism developed here\re{abelianhiggs}.

While we do not yet have all the answers we need to fully understand what
changes the time dependence of the rate will bring, we may speculate. We have
considered thermal activation here rather than under the barrier tunnelling.
This makes the range of applicability of our calculation to inflationary models
somewhat suspect. The reason for this is that once inflation sets in, the
temperature of the heat bath will decrease rapidly, turning the problem into a
zero temperature one. Even in this case, however, we should expect the rate to
be time dependent. The basic change in our calculation is that the paths used
to compute the propagators requied to evolve the density matrix in time will be
different. Essentially, one must do a WKB approximation of the
propagators\re{danme}. Again, though, we would expect the rate to start at zero
and approach zero asymptotically at large times. This implies the rate must
peak at some time, just as our current calculation. Our work can therfore be
used to provide some hints as to what might occur in the zero
temperature case.

In the case of Guth's original inflationary scenario, it seems to us that the
time dependence of the rate just exarcerbates the problems that led to its
downfall. Recall that the problem had to do with the fact that the nucleation
rate had to be small enough to keep the system in the false vacuum long enough
to achieve a sufficient amount of inflation yet large enough so that the phase
transition would be completed (these requirements can be quantified more
explicitly\re{guthwein}). Our rate will start off being small and then grow to
a maximum and finally tail off to zero. Thus, just when a large rate is needed
to complete the phase transition, the rate is getting small. There may be ways
to avoid this. For example, if the rate starts off small enough, there may be
enough time before the rate peaks to achieve the requisite $60$ e-folds of
inflation. The rate could still be growing past this time in a way that would
allow the new phase to percolate.

The situation in extended inflationary models is somewhat trickier to assess,
since the nucleation rate is already time dependent in most of these models due
to the time evolution of the Jordan-Brans-Dicke field in them\re{us}.

In the case of the sphaleron, the question of whether the sphaleron
interactions are in thermal equilibrium (which is crucial in terms of
determining whether a baryon asymmetry can be generated by these interactions),
becomes more difficult to assess due to the time dependence of the rate. One
could imagine that the rate of these interactions
decreased sufficiently quickly so as to allow them to drop out of local thermal
equilibrium thus allowing a net $B$ asymmetry to be generated. However, the
answer to these and other questions will only be found when the field theoretic
generalization of our calculation is completed.

It seems clear then that a {\it real time} calculation of the rate of thermal
activation of a metastable state will always yield a time dependent rate. This
time dependence is completely missed in the standard equilibrium calculations
due to a choice of initial state that is not realized in the situations
these calculations are usually applied to. Furthermore, there are now
experimental results that support our arguments and that {\it cannot} be
explained by the usual homogeneous nucleation theory. We believe that our
methods will have wide applicability to a variety of problems in this branch of
physics, not least of which is the understanding of some very important facets
of early universe physics.

\vspace{18pt}
\centerline{\bf Acknowledgements}
D.B. would like to thank David Jasnow and Walter Goldburg for useful
discussions. D.B. and D-S.L were supported in part by NSF grant $\#$
PHY-8921311
as well as a Mellon Pre-Doctoral Fellowship Award (D-S.L). R. H., J.S. and A.S.
were supported in part by DOE grant $\#$ DE-FG02-91ER40682.The work of J.S. is
also supported in part by the Portuguese JNICT, under CIENCIA grant $\#$
BD/374/90-RM.

\vspace{36pt}

\centerline{\bf References}
\frenchspacing

\begin{enumerate}

\item\label{sphaleron} For a good review, see ``Anomalous Fermion Number
Non-Conservation'' by M.E. Shaposhnikov, CERN preprint CERN-TH.6304/91 (1992).

\item\label{baryogenesis} See e.g. ``Non-GUT Baryogenesis'' by A. D. Dolgov
{\em Phys. Rep.} {\bf 222}, 309 (1992).

\item\label{lastein}D. La and P. J. Steinhardt, {\em Phys. Rev. Lett.}
        {\bf 62}, 376 (1989).

\item\label{guth}A.H. Guth, {\em Phys. Rev.} {\bf D 23}, 347 (1981).

\item\label{langer}J.S. Langer, {\em Ann. Phys.} {\bf 41}, 108 (1967);
{\em ibid} {\bf 54} 258 (1969).

\item\label{affleck}Ian Affleck, {\em Phys. Rev. Lett.} {\bf 46}, 388 (1982).

\item\label{feynman} `` Statistical Mechanics'' by R.P. Feynman,
W.A. Benjamin Inc., Reading, Mass. (1972).

\item\label{barton}G. Barton, {\em Ann. Phys.} {\bf 166}, 322 (1986).

\item\label{mingoldburg} K.Y. Min and W. Goldburg, ``Dependence of Nucleation
on Shear Induced Initial Conditions'' (submitted to {\em Phys. Rev. Lett.})
(1993).

\item\label{dc}``Real Time Analysis of Thermal Activation via Sphaleron
Transitions'' by D. Boyanovsky and C. Arag\~{a}o de Carvalho, University of
Pittsburgh Preprint PITT-93-5 (1993).

\item\label{abelianhiggs}D. Boyanovsky, R. Holman, D-S. Lee, J.P. Silva and
A. Singh,(work in progress)\\ (1993).

\item\label{danme}D. Boyanovsky, R. Holman and R. Willey, {\em Nucl. Phys.}
{\bf B376}, 599 (1992).

\item\label{guthwein}A.H. Guth and E.W. Weinberg, {\em Nucl. Phys.} {\bf B212},
321 (1982).

\item\label{us} R. Holman, E.W. Kolb, S.L. Vadas and Y. Wang, {\em Phys. Lett.}
{\bf 250B}, 24 (1990).

\end{enumerate}
\newpage
\nonfrenchspacing
\centerline{\bf Figure Captions}

\vspace{36pt}

Figure 1: The potential for a 1 dimensional metastable system

Figure 2: The thermal activation rate $\Gamma(t)$ as a function of $t$ at fixed
temperature. Time is measured in units of $\Omega^{-1}$, while the rate is
measured in arbitrary units.

Figure 3: Plots of $\Gamma(t)$ at different temperatures $T_1$, $T_2$, $T_3$,
with $T_1>T_2>T_3$. The $T_i$ are taken so that the system can be thought of as
metastable.

\end{document}